\documentstyle[sprocl]{article}

\def\be{\begin{equation}}
\def\ee{\end{equation}}
\def\bea{\begin{eqnarray}}
\def\eea{\end{eqnarray}}
\input epsf

\begin{document}

\title{MONOPOLES IN QCD}

\author{H.~ICHIE and H.~SUGANUMA}

\address{Research Center for Nuclear Physics, Osaka University
\\E-mail: ichie@rcnp.osaka-u.ac.jp
}


\maketitle
\abstracts{ 
Based on the dual Higgs picture for confinement,
we study 
monopole properties in the maximally abelian (MA) gauge
using the lattice QCD.
The monopole carries a large fluctuation of the gluon field
and provides a large abelian action in abelian projected QCD (AP-QCD).
We find large cancellation between abelian part and off-diagonal 
part of the action density around the monopole, which ensures the 
appearance of monopoles.
The monopole in the MA gauge can be regarded as the collective mode of 
the large gluon fluctuation   concentrated into the abelian sector, and 
would be relevant degrees of freedom for confinement in AP-QCD. 
}

In the 't~Hooft abelian gauge, QCD is reduced into an abelian gauge 
theory with magnetic monopoles\cite{thooft},  
and color confinement can be understood 
by the dual Meissner effect with monopole condensation.
In relation to the dual Higgs picture, recent
lattice QCD simulations show strong evidence of 
abelian dominance and  monopole dominance for confinement in the 
maximally abelian  (MA) gauge.
Abelian projected QCD (AP-QCD) is defined by neglecting
off-diagonal gluon components in the MA gauge, and it almost 
reproduces the essence of nonperturbative features of QCD.

We investigate the theoretical structure of AP-QCD by comparing
with QCD and QED as shown in Fig.1.
As for confinement properties, AP-QCD resembles the original QCD, while
QED does not provide  confinement phenomena.
The difference between QCD and QED is existence of off-diagonal 
components of the gauge field, which leads to the asymptotic freedom and 
strong-coupling nature in the infrared region of QCD.
However, AP-QCD has an abelian gauge symmetry and resembles QED rather 
than QCD on the symmetry.
Then, why does AP-QCD provide confinement without off-diagonal gluons ?
The obvious difference between AP-QCD and QED is  existence of  
monopoles.
As a natural conjecture, the monopole in AP-QCD plays an alternative role of 
off-diagonal gluons and is responsible for confinement phenomena.
Then, the monopole would be closely related to off-diagonal gluons.

Next, we consider  monopole properties in terms of the action.
In the static frame of the monopole 
with the magnetic charge $g$, 
a spherical `magnetic field' is created around the monopole 
in the abelian sector of QCD as 
%\begin{eqnarray}
${\bf H}(r) =  \frac{g}{4\pi r^3} {\bf r}$
%= \frac{{\bf r}}{e r^3}
%\end{eqnarray}
with ${\bf H}_i\equiv \epsilon _{ijk} \partial_j A^3_k$. 
%and the Dirac condition $eg=4\pi$ is used.
Then, the monopole inevitably accompanies 
a large fluctuation of the abelian gluon component $A^3_\mu$ around it. 
For the abelian part $S_{\rm Abel}\equiv -\frac14 \int d^4x
(\partial_\mu A_\nu^3-\partial_\nu A_\mu^3)^2$ of the QCD action, 
the electro-magnetic energy
created around the monopole 
is estimated as 
%\begin{eqnarray}
$E(a) = \int_a^\infty d^3 x \frac12 {\bf H}(r)^2 = 
\frac{g^2}{ 8 \pi a},$
%= \frac{2 \pi}{e^2 a},
%\label{monopolemass1}
%\end{eqnarray}
where $a$ is an ultraviolet cutoff like a lattice mesh.
As the ``mesh'' $a$ goes to 0, 
the monopole provides a large fluctuation of 
$S_{\rm Abel}$, 
and hence the monopole seems difficult to appear if the abelian 
gauge theory is controlled by $S_{\rm Abel}$.
This is the reason why QED does not have the point-like Dirac monopole.
Then, why can the monopole appear in the abelian sector of QCD ?
To answer it, let us consider the division of the total QCD action 
$S_{\rm QCD}$ into the abelian part $S_{\rm Abel}$ and 
the remaining part $S_{\rm off}\equiv S_{\rm QCD}-S_{\rm Abel}$, which 
is contribution from the off-diagonal gluon component. 
Unlike $S_{\rm QCD}$ and $S_{\rm Abel}$, 
$S_{\rm off}$ is not positive definite and can take a negative value
in the Euclidean metric.
Then, around the monopole, 
the abelian action $S_{\rm Abel}$ should be 
partially canceled by the remaining  off-diagonal contribution $S_{\rm off}$
such that the total QCD action $S_{\rm QCD}$ around the monopole 
does not become extremely large. 
Thus, we expect large off-diagonal gluon components 
around the monopole for  cancellation with the
large field fluctuation of the abelian part.
Based on this analytical consideration, we study
action densities around monopoles in the MA gauge using the lattice QCD.

On the SU(2) lattice, we measure the action densities 
$\bar S_{\rm SU(2)}$, $\bar S_{\rm Abel}$ and $\bar S_{\rm off}$,
which are the SU(2), the abelian and the off-diagonal parts, respectively. 
We show 
the probability distribution 
of 
$\bar S_{\rm SU(2)}$, $\bar S_{\rm Abel}$ and $\bar S_{\rm off}$ in Figs.2 and 3.
(To be exact, $\bar S$ is the averaged value over the neighboring links 
around a dual link.)
For total distribution on the whole lattice,  most 
$\bar S_{\rm off}$ is positive, and both $\bar S_{\rm Abel}$ and 
$\bar S_{\rm off}$ tend to take smaller values than $\bar S_{\rm SU(2)}$
owing to 
$\bar S_{\rm SU(2)} = \bar S_{\rm Abel} +  \bar S_{\rm off}$. 
Around the monopole, however, the off-diagonal part $\bar S_{\rm off}$ of 
the action density tends to take a large negative value, and  
$ \bar S_{\rm off}$ strongly cancels with the large abelian action 
density  $\bar S_{\rm Abel}$ so as to keep the total SU(2) action small.
Thus, monopoles can appear in the abelian sector in QCD without large cost 
of the QCD action due to large cancellation between the abelian 
action $ S_{\rm Abel}$ and the off-diagonal part $ S_{\rm off}$ 
of the action.

In AP-QCD, the monopole provides a large abelian field fluctuation.
For confinement, this large abelian field fluctuation originated from 
monopoles
would be important in AP-QCD in terms of the area-law 
reduction of the abelian Wilson loop,
which leads to the abelian string tension.
In conclusion, the monopole in the MA gauge can be regarded as the 
collective mode of the large gluon fluctuation
concentrated into the abelian sector, and would be relevant degrees of 
freedom for 
confinement in AP-QCD.

%One of authors (H.I.) is supported by 
%Research Fellowships of the Japan Society for the 
%Promotion of Science for Young Scientists.

\hspace{5mm}

%
%----Figures
%

\hspace{2cm}

\noindent {\bf \large Figure Captions}

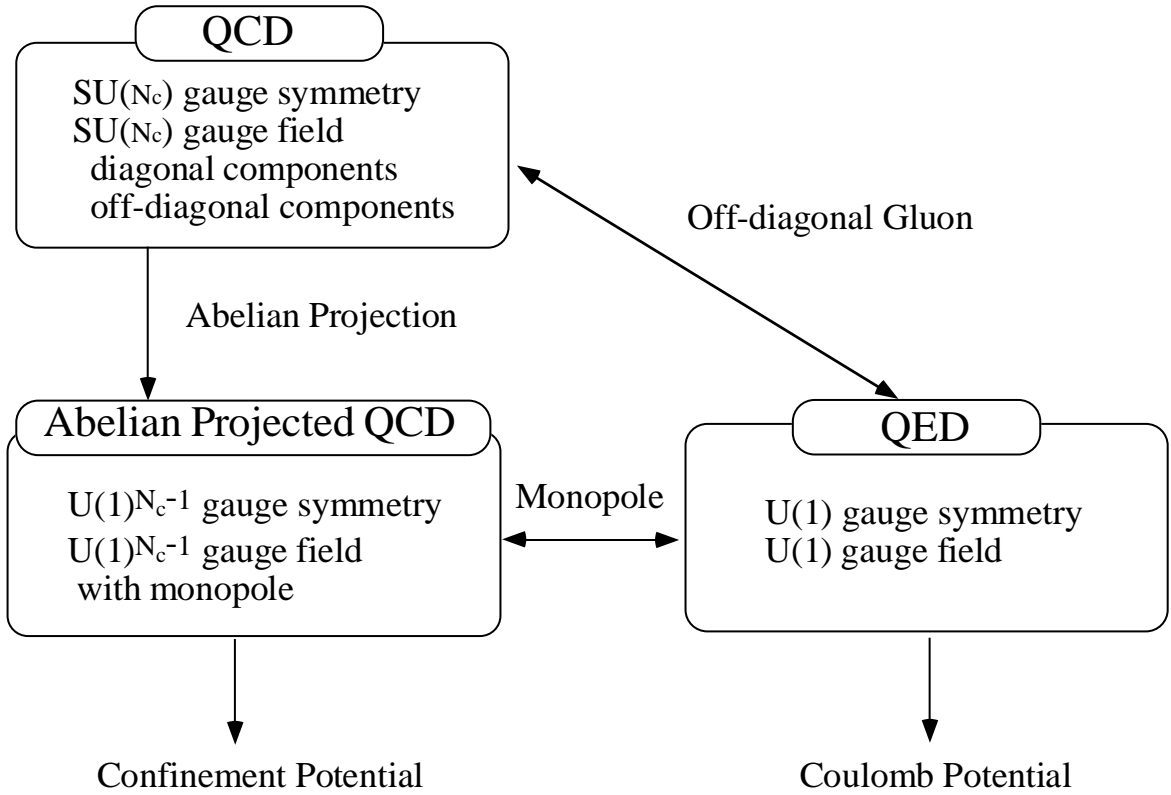
\noindent {\bf Fig.1}
Comparison among QCD,  abelian projected QCD (AP-QCD) and QED in terms of 
the gauge symmetry and fundamental degrees of freedom. 

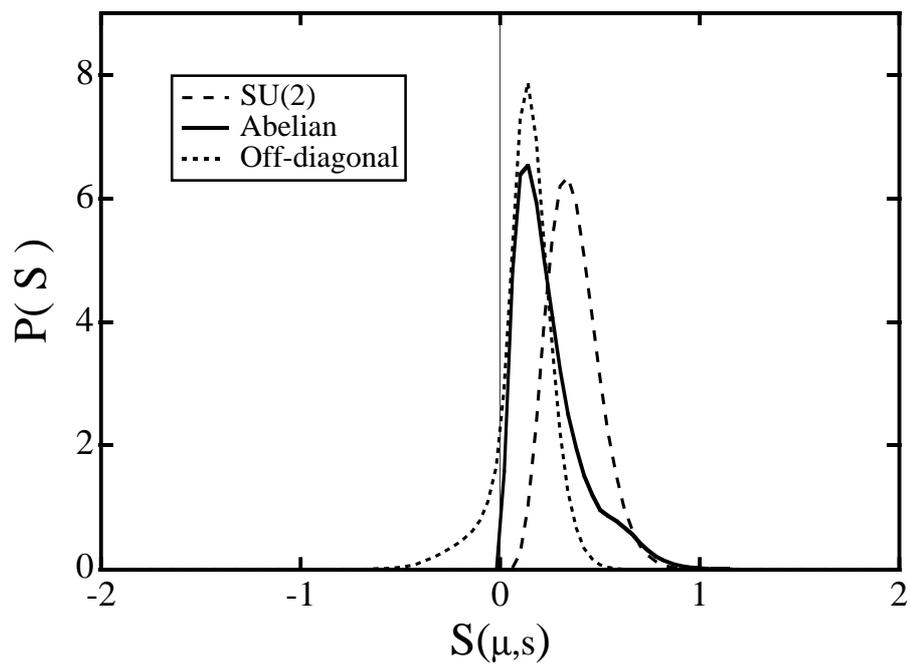
\noindent {\bf Fig.2}
Probability distributions $P(\bar S)$ of the 
SU(2) action density $\bar S_{\rm SU(2)}$(dashed curve), abelian action density
$\bar S_{\rm Abel}$ (solid curve) and off-diagonal part $\bar S_{\rm off}$
(dotted curve) in the MA gauge at $\beta =2.4$ on $16^4$ lattice.

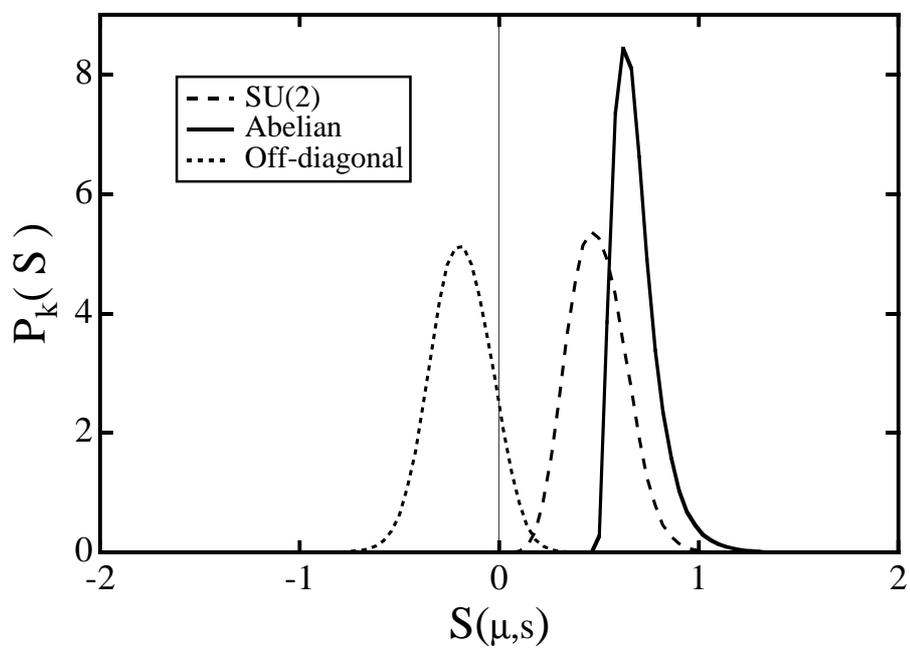
\noindent {\bf Fig.3}
The probability distribution $P_k(\bar S)$ 
{\bf around the monopole} for $\bar S_{\rm SU(2)}$, $\bar S_{\rm Abel}$ 
and $\bar S_{\rm off}$
in the MA gauge at $\beta =2.4$ on $16^4$ lattice.
The meaning of curves is same as Fig.2.
Large cancellation between $\bar S_{\rm Abel}$ and  
$\bar S_{\rm off}$ is found.


\section*{Reference}
\begin{thebibliography}{99}
     \bibitem{thooft} 
G.~'t~Hooft, Nucl.~Phys.~{\bf B190} (1981) 455.
\end{thebibliography}
\end{document}